\documentclass[aps,prl,showpacs,floatfix,twocolumn,superscriptaddress]{revtex4}
\usepackage{graphicx,amsmath,amssymb}
\usepackage{epsf,color}
\usepackage{latexsym}

\begin{document}
\title{Optimization of quantum interferometric metrological sensors \\
  in the presence of photon loss}  
\author{Tae-Woo Lee}
\affiliation{Center for
  Computation and Technology, Louisiana State University, Baton Rouge,
  Louisiana 70803, USA}
\author{Sean D. Huver}
\affiliation{Hearne Institute for Theoretical Physics, Department of Physics
  and Astronomy, Louisiana State University, Baton Rouge, Louisiana 70803, USA}  
\author{Hwang Lee} 
\affiliation{Hearne Institute for Theoretical Physics, Department of Physics
  and Astronomy, Louisiana State University,
 Baton Rouge, Louisiana 70803, USA}
\author{Lev Kaplan}
\affiliation{Department of Physics, Tulane University, New Orleans, Louisiana 70118, USA}
\author{Steven B. McCracken} 
\affiliation{Hearne Institute for Theoretical Physics, Department of Physics
  and Astronomy, Louisiana State University,
 Baton Rouge, Louisiana 70803, USA}
\author{Changjun Min}
\affiliation{Center for
  Computation and Technology, Louisiana State University, Baton Rouge,
  Louisiana 70803, USA}
\author{Dmitry B. Uskov} 
\affiliation{Department of Physics, Tulane University, New Orleans, Louisiana 70118, USA}
\author{Christoph F. Wildfeuer} 
\affiliation{Hearne Institute for Theoretical Physics, Department of Physics
  and Astronomy, Louisiana State University,
 Baton Rouge, Louisiana 70803, USA}
\author{Georgios Veronis}
\affiliation{Center for
  Computation and Technology, Louisiana State University, Baton Rouge,
  Louisiana 70803, USA}
\affiliation{Department of Electrical and Computer Engineering,Louisiana State University,
 Baton Rouge, Louisiana 70803, USA}
\author{Jonathan P. Dowling}
\affiliation{Hearne Institute for Theoretical Physics, Department of Physics
  and Astronomy, Louisiana State University,
 Baton Rouge, Louisiana 70803, USA}

\begin{abstract}
We optimize two-mode, entangled, number states of light in the presence of loss in order to maximize the extraction 
of the available phase information in an interferometer. Our approach optimizes over the entire available input Hilbert 
space with no constraints, other than fixed total initial photon number. We optimize to maximize the Fisher information, 
which is equivalent to minimizing the phase uncertainty. We find that in the limit of zero loss the optimal state is the 
so-called N00N state, for small loss, the optimal state gradually deviates from the N00N state, and in the 
limit of large loss the optimal state converges to a generalized two-mode coherent state, with a finite total number 
of photons. The results provide a general protocol for optimizing the performance of a quantum optical 
interferometer in the presence of photon loss, with applications to quantum imaging, metrology, sensing, and 
information processing.\end{abstract}

\maketitle

Quantum states of light play an important role in applications including metrology, imaging, sensing, and 
quantum information processing \cite{Hwang}. In quantum interferometry, entangled states of light, such as the
maximally path-entangled N00N states, replace conventional laser light to achieve a sensitivity below the shot-noise 
limit, even reaching the Heisenberg limit, and a resolution well below the Rayleigh diffraction limit \cite{Boto}. For an 
overview of quantum metrology applications see, for example, Ref.~\cite{Hwang}. However, for real-world applications, diffraction, scattering, and absorption of quantum states of light need to be taken into account. Recently it has been 
shown that many quantum-enhanced metrology schemes using N00N states perform poorly when a considerable 
amount of loss is present \cite{gilbert08,Rubin,Rodecap}. However, our team has also discovered a new class of 
entangled number states, which are more resilient to loss \cite{huver08}. These so-called $M \& M^\prime$ states 
still outperform classical light sources under a moderate 3 dB of loss. 

In this work, we systematize the numerical search for optimal quantum states in a two-mode interferometer in the 
presence of loss. We employ the Fisher information to obtain the phase sensitivity of the interferometer. An exhaustive review and application of the Fisher information concept to the sensitivity of a March-Zehnder interferometer, particularly in the zero loss case, has been presented in the recent work by Durkin and Dowling \cite{Durkin}. The chief utility of the Fisher information approach is that it provides a bound on the phase sensitivity, even in the absence of a fully specified detection scheme, and is now widely adopted in studies of interferometer sensitivity. Such numerical optimization has been previously carried out in the absence of loss, and with loss 
over a restricted class of input states \cite{uys,Doner}. Here, we provide a completely general optimization scheme 
that is applied to the two-mode interferometer, but also has application to the optimization of linear optical systems 
for quantum linear optical information processing \cite{Kok07,Dmitry09}. 

Using this scheme, we first recover the well-known fact that N00N states are optimal in the absence of loss \cite{Hwang}.
For large loss, the optimal states belong to a class of two-mode coherent states 
with finite total photon number. The optimization procedure yields the optimal Fisher information -- and hence the minimal phase 
uncertainty -- for every level of loss. The validity of our numerical optimization
is verified using several methods, including genetic algorithms and simulated annealing, and the close agreement among these methods provides evidence that we are indeed finding the global optimum. 

In quantum optics,  photon loss is typically modeled by a beam splitter that routes 
\begin{figure}[htb]
\includegraphics[scale=0.581]{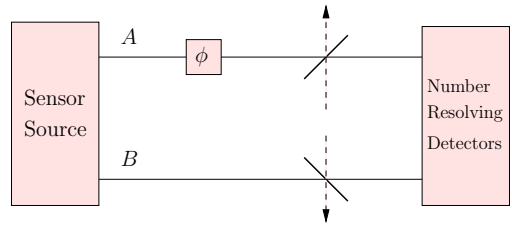}
\caption{(Color online) Abstract interferometer condensing the input state plus the first beam splitter into the 
  first box, followed by two propagating modes with loss modeled by additional beam splitters. The box on the 
  right includes a beam splitter and the photon-number resolving detectors.\label{interferometerwithloss}}
\end{figure}
photons out of the interferometer \cite{loudon00}. In implementing this model, we first enlarge the Hilbert space 
to include modes representing the scattered photons and then, after the scattering, trace out these modes.  
Here we extend the beam splitter model for photon loss to two propagating modes that represent the two paths 
in the optical interferometer (Fig.~\ref{interferometerwithloss}). 

For an interferometer with two input ports A and B as depicted in Fig.~\ref{interferometerwithloss}, 
an arbitrary pure-state 
input with $N$ 
photons can be written as  $|\psi \rangle_{\rm input} = \sum_{k=0}^{N} c_k |N-k,k\rangle_{A,B}$
where $c_k$ are the input amplitudes to be optimized. The ket 
$|N-k,k\rangle_{A,B}$ is a basis state in which 
$N-k$ and $k$ photons are in mode-A (detection arm) and mode-B (control arm), respectively.
Such a quantum state resides in an $N+1$ dimensional Hilbert space. The transformation of the quantum state 
by any passive lossless optical elements, such as beam splitters, phase shifters, and mirrors, can then be 
described by an $(N+1)\times(N+1)$ unitary matrix. 

When the propagation channels suffer from photon loss, we need to consider the total density matrix that 
includes all the scattered photon modes. Then we obtain the reduced density matrix for the two 
interferometer modes by tracing out these additional modes \cite{Glauber}.
This reduced density matrix is associated with a much larger Hilbert space of dimension $(N+1)(N+2)/2$, which includes
all states with a total of $N, N-1, ..., 0$ photons in the two interferometer modes.

For the N00N state, combined with its specific detection scheme, the density-matrix description of photon loss 
may be avoided and the state-vector approach can be adopted \cite{Rubin,gilbert08,gao08}. Previously, we have 
used the density-matrix approach to lossy interferometers for particular input states of light (namely, the 
$M \& M^\prime$ states \cite{huver08}). Our method here, however, 
applies to any input state with a fixed number of photons. Thus, it allows optimization of the input state in the presence of an arbitrary amount of propagation loss in the 
two arms of an interferometer. 

With this scheme, the pure state input is now described as an $(N+1)\times(N+1)$ density matrix. After passing 
through the two beam splitters --- representing the two lossy channels in the two arms of the interferometer --- the 
density matrix for the two main modes now consists of $N+1$ different blocks. Each block represents a given number of photons lost.  The quantum state of light ends up in a mixed state associated with an $N+1-L$ dimensional Hilbert 
space, if a total of $L$ photons are lost.

For the arbitrary input state $|\psi\rangle_{\rm input}$ presented above, we find that photon loss can be 
described by a transformation to a reduced density matrix whose matrix elements are simply given by
\begin{eqnarray}
\rho_{N,L,i,j} = \sum_{m=0}^{L} 
c_{i+m} \, c_{j+m}^* \,
A_{N,L,i,m} \, B_{N,L,j,m}^* \,,
\label{rdm-elements}
\end{eqnarray}
\noindent
where 
\begin{eqnarray}
\lefteqn{A_{N,L,k,m} = B_{N,L,k,m}=} \nonumber \\
     &     \sqrt{C^{N-k-m}_{N-L-k}}~t^{N-L-k} ~r^{L-m}
\sqrt{C^{k+m}_{k}}t^{\prime k} r^{\prime m} \,.
\label{coefficients}
\end{eqnarray}
\noindent
Here, $t$, $r$ and $t^\prime$, $r^\prime$ are transmission and reflection coefficients for the fictitious beam splitters 
in the upper path and the lower path, respectively, and $C$ is the number of combinations,
$C^n_k = {n \choose k}$. Note that for a given number of photons $N$, the $L$ value labels the block 
of the reduced density matrix, and $i$, $j$ specify the matrix element inside 
that block. 

Now we compare a classical interferometer with the optimized quantum state approach. Consider a single-mode 
coherent state  $|\alpha\rangle=\mathrm{e}^{-\frac{|\alpha|^2}{2}}\sum_{k=0}^\infty\frac{\alpha^k}{\sqrt{k!}}|k\rangle $ 
as the input state to a Mach-Zehnder interferometer (MZI) \cite{Glauber}. The first beam splitter partitions the state 
between two modes. In the first mode, the state acquires a phase shift $\phi$ and undergoes a loss of intensity by a 
factor of $|t|^2$. Then, the two beams are redirected to the second beam splitter, and photons are detected in each output 
port. The Fisher information, normalized to the average number of photons $\bar{n}=|\alpha|^2$, is 
$F/\bar{n}=(4|t|^2\sin^2\theta)/(1+|t|^2\tan^2\theta)$, where $\theta$ describes the angle of $\sigma_y$ rotation by 
the first beam splitter. This equation provides the Cramer-Rao asymptotic accuracy of measurement of the unknown 
phase shift $\phi$ using a classical scheme. We single out two cases. First, we assume that the first beam splitter has 
a fixed 50-50 ratio. Then in the limit of large loss $t \rightarrow 0$ we have
\begin{equation}\label{limit1}
  F/\bar{n}=\frac{2|t|^2}{1+|t|^2}\stackrel{|t|\rightarrow
  0}{\rightarrow}2|t|^2\,.
\end{equation}
Fisher information can be increased by optimizing the first beam splitter to compensate for the loss. In this case we 
obtain
\begin{equation}\label{limit2}
  F/\bar{n}=\frac{4|t|^2}{(1+|t|)^2}\stackrel{|t|\rightarrow
  0}{\rightarrow}4|t|^2\,.
\end{equation}

The optimal quantum input state is now obtained numerically.
First, a forward problem solver is developed using a density matrix approach. An input state is written as a density 
matrix. Phase shifts during photon propagation are taken into account by  operating with $e^{i \phi \widehat{n}_A}$ 
on this density matrix. Next, photon losses are applied using Eqs. (\ref{rdm-elements}) and (\ref{coefficients}), 
producing a reduced density matrix of dimension $(N+2)(N+1)/2$. In the last step of the forward problem solver, 
the minimum detectable phase sensitivity $\delta\phi$ is computed from the final density matrix. The phase detection 
is modeled by a final 50-50 beam splitter followed by two number resolving photodetectors. 
The joint probability of simultaneously detecting $m_1$ photons at the first photodetector and $m_2$ 
photons at the second photodetector is computed as $P_{m} = \sum_{i=1}^{(N+2)(N+1)/2} \hat{U}_{\text{bs}_{m,i}} 
\sum_{j=1}^{(N+2)(N+1)/2} \hat{\rho}_{\text{out}_{i,j}} \hat{U}_{\text{bs}_{j,m}}^{\dag}$, where the label $m$ represents 
a pair of numbers ($m_1$, $m_2$). Here, $\hat{U}_{\text{bs}}$ is a unitary transformation representing a 50-50 beam 
splitter and $\hat{\rho}_{\text{out}}$ is the final density matrix obtained after loss. Then, phase sensitivities are 
estimated from the Fisher information, $F$, for a single measurement, $\delta\phi = 1/\sqrt{F}$, where $F = \sum_{m = 1}^{(N+2)(N+1)/2} P_m \left(\partial \ln P_{m} / \partial \phi \right)^2$ \cite{uys}. We note that, in all of our calculations, we assume a large flux of entangled states is to be used, and we normalize our results by this flux.

We optimize the system to find the minimum detectable phase sensitivity, given fixed losses in the detection and 
control arms. For this, a genetic global optimization algorithm is applied to the forward problem solver. The parameters 
to be optimized are the complex coefficients $c_k$; the optimal sensitivity is necessarily $\phi$-independent since a change in $\phi$ can be absorbed into the relative phases of $c_k$. During the numerical computation of $F$, we observe that the landscape of $F$ in the optimization parameter space possesses several local maxima contrary to the convex $\tilde{F}_Q$ used by Doner {\em et al.} \cite{Doner}.

\begin{figure}[htb]
\includegraphics[scale=0.49]{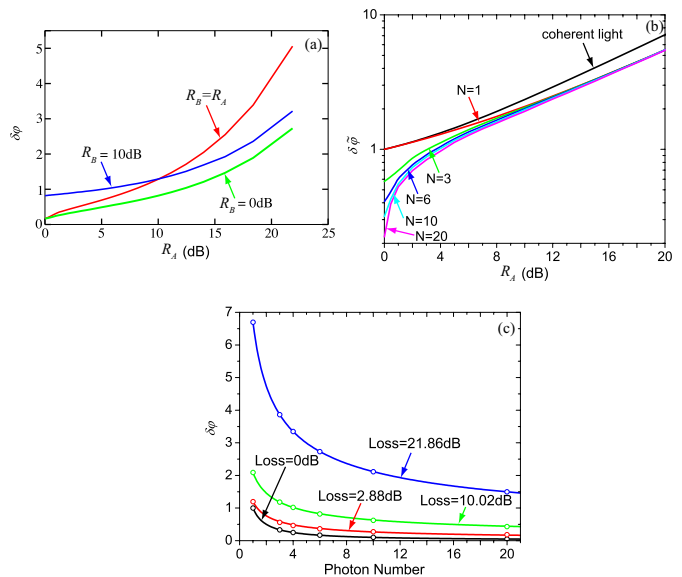}
\caption{(Color online) 
Minimum detectable phase sensitivities calculated from the normalized Fisher information: (a) As a function of detector arm loss ($R_A$) with three different losses in the control arm ($R_B$) for $N = 6$. (b) As a function of $R_A$ for $R_B = 0$ in log scale. (c) As a function of input photon number $N$ for $R_B = 0$ with fixed $R_A$. Lines represent the results of curve fitting using a functional form $1/N^x$. In the absence of loss, the Heisenberg limit, $x = 1$, is obtained. For high loss, $x$ tends toward $0.5$, approaching the shot noise scaling. \label{del_phi}
}
\end{figure}

The results of numerical optimization of $\delta\phi$ are presented in Fig.~\ref{del_phi}. We denote the losses in 
dB in the detection and control arms as $R_A$ and $R_B$, respectively. First, in Fig.~\ref{del_phi}(a)
we investigate the 
overall influence of loss in the control arm.  One set of simulations is conducted with equal losses in the detection 
and control arms ($R_A = R_B$). We also consider fixed 10 dB loss and fixed 0 dB loss in the control arm ($R_B = 10$ dB, $R_B = 0$ dB) as loss in the detection arm is varied. In all cases, $N=6$ is assumed.
We consistently find that an increase in $R_B$ results in higher $\delta\phi$. Thus, one can expect the best phase sensitivity to be achieved with the 
smallest possible loss in the control arm. 

In Fig.~\ref{del_phi}(b), the numerically optimized phase sensitivity is presented as a function of $R_A$ for 
$N=1$, $3$, $6$, $10$, and $20$. The classical (coherent light) baseline of Eq.~(\ref{limit1}) is also shown. 
To compare the quantum results for different $N$ with each other and with the classical case, we rescale the phase sensitivity of the pure quantum state by normalizing
the Fisher information similarly to Eqs.~(\ref{limit1}) and (\ref{limit2}): $\delta\tilde{\phi} = 1/\sqrt{F/N}=\delta\phi \sqrt{N}$. Since 
$\delta\tilde{\phi}$ is obtained from the Fisher information per single photon, it is also the measure of the synergically-enhanced phase sensitivity per single photon with $N$ photons acting together. For pure quantum states,   Fig.~\ref{del_phi}(b) shows 
that larger $N$ produces smaller $\delta\tilde{\phi}$ for any given amount of loss. 
The $N$-dependence of $\delta\tilde{\phi}$ is greatest at 
$R_A = 0$ dB, and weakest in the limit of extremely high loss, where the lines merge together. Coherent light does not show enhancement with $N$ at any level of loss, and at every level of loss, coherent light exhibits worse performance in phase sensitivity compared to entangled quantum states.

One interesting observation is that $\delta\tilde{\phi}$ in the extremely high loss region ($R_A > \: \sim$ 16 dB) 
becomes $N$-independent even when using optimally entangled quantum states. In other words, the optimal phase sensitivity $\delta \phi$ given by the optimal quantum state becomes
proportional to $N^{-1/2}$ in this high-loss regime, i.e., its scaling with $N$ in this regime is the same as for coherent light governed by the shot noise limit. However, despite the same scaling with $N$, the phase sensitivity $\delta\phi$ 
is still better with entangled quantum states than with coherent light. This can be explained by the optimal preparation of the initial state. As we will see later, the probability amplitudes $c_k$ are distributed asymmetrically to 
generate the smallest possible $\delta\phi$ for nonzero loss, while coherent light always enters the system through 50-50 beam 
splitter, i.e., it is symmetrical between the control and detection arms, Eq.~(\ref{limit1}). With coherent light, a similar 
improvement can be achieved by adjusting (or optimizing) the first beam splitter, resulting in Eq.~(\ref{limit2}). In the latter 
case, $\delta\tilde{\phi}$ of coherent light becomes identical to that of the pure quantum state with $N = 1$. However, we emphasize that, when losses are not too high, the phase sensitivity of $N > 1$ pure quantum states is always better than that of
coherent light, even with an optimized first beam splitter.

\begin{figure}[ht]
\includegraphics[scale=0.48]{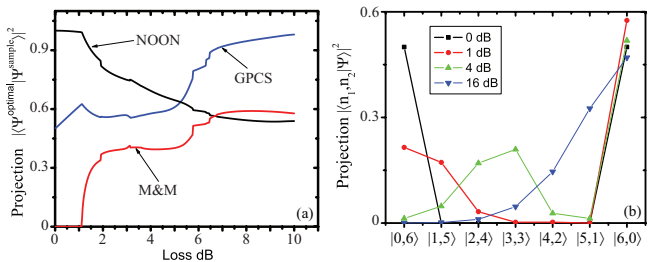}
\caption{(Color online) (a) Projection of the optimal state on N00N, $M \& M^\prime$, and GPCS as a function of $R_A$, for $N=4$.  
(b) The optimal input state composition. The vertical axis shows coefficients $|c_k|^2$ of the optimally entangled input state. 
Both figures clearly demonstrate that the optimal 
state changes from N00N-type to GPCS-type as loss increases, with the crossover occurring at approximately 5 dB loss. \label{c_amp}
}
\end{figure}

Figure \ref{del_phi}(c) shows the optimal phase sensitivity as a function of photon number $N$ for given fixed $R_A$ values, with
$R_B=0$. Here $\delta\phi$ is shown for each $N$, with circles indicating
$N =$ 1, 3, 4, 6, 10, and 20. 
The lines are drawn by curve-fitting to a power law, $\delta \phi \sim 1/N^x$. From this, we find that $\delta\phi$ is well 
represented by $1/N$, $1/N^{0.68}$, $1/N^{0.517}$, and $1/N^{0.5}$ for $R_A = 0$ dB, 2.88 dB, 10.02 dB, and 
21.86 dB, respectively. 
This result provides an overall view  of how the phase sensitivity changes from the Heisenberg limit to the shot noise 
limit with increased loss, i.e., it follows a power law with $1/N^x$ where $1/2 \leqslant x \leqslant 1$.

To characterize the optimal state, we use three classes of well-defined states: N00N, $M\&M^\prime$, and a two-mode SU(2) coherent state, often called the Generalized Perelomov Coherent State (GPCS). GPCS is defined as $|GPCS\rangle=(N!)^{-1/2} [\hat{a}_1^\dagger e^{i\beta}\cos{\alpha} - \hat{a}_2^\dagger e^{-i\beta} \sin{\alpha}]^N
|0\rangle$, where $\hat{a}_1^\dagger$ and $\hat{a}_2^\dagger$ are creation operators in the two modes,
and $\alpha$ and $\beta$ are two real parameters \cite{Perelomov}. In particular, $M\&M^\prime$ is the first class of path entangled states shown analytically to have robustness to photon loss. It is interesting to see how the true optimal state may differ from the $M\&M^\prime$ state in lossy environments.
Characteristics of the optimal state are presented in Fig.~\ref{c_amp} for a fixed $R_B=0$ dB. The similarity of the optimal states with each of these three benchmark states is measured by the squared overlap between the optimal state and the benchmark state. 
The results show that the optimal 
state is closest to the N00N state for low loss. As loss increases, the N00N state portion gradually decreases and the optimal state becomes closer to GPCS than to N00N at around 5 dB of loss. 
The degree of similarity between the optimal state and the $M \& M^\prime$ state is rather low for loss smaller than 6 dB. For every value of loss, GPCS is higher than $M \& M^\prime$. Figure \ref{c_amp}(b) shows how the input amplitudes of  the 
optimal state are arranged for different loss levels. In the lossless case, we have the N00N state. 
As loss increases, the optimal state is reshuffled and acquires an asymmetric shape. 
This serves as critical information for achieving a highly sensitive interferometric system. 
Based on the results shown, it is obvious that generating such optimal input states should be the first consideration in the 
development of an interferometric sensor using entangled photons.

In summary, we have performed unconstrained optimization of a lossy two-mode interferometer. 
We conclude that input N00N states are optimal for nearly zero loss [1], and that finite-photon number two mode 
coherent states are optimal --- with shot-noise sensitivity ---  for large loss. 
Our results suggest that, if sensitivity is the only metric of success, ordinary coherent input state 
interferometry is best for high loss. This leaves open super-sensitive schemes employing squeezed light 
at the detector \cite{Caves} or super-resolving schemes employing photon number resolving detectors \cite{Gao09}.

\begin{acknowledgments}
We would like to acknowledge the Army Research Office, the Boeing Corporation, 
the Defense Advanced Research Project Agency, the Department of Energy,
the Foundational Questions and Extreme Science Institute, 
the Intelligence Advanced Research Projects Activity, and
the Northrop-Grumman Corporation. This work was also supported in part by 
NSF KITP Grant PHY05 - 51164 and NSF Grant 0545390. 
Computer resources are provided in part by Louisiana Optical Network Initiative, or LONI.
\end{acknowledgments}


\end{document}